\documentclass[conference]{IEEEtran}

\usepackage{graphicx,mathptmx,amsmath,amsfonts,amsthm,color,comment}
\usepackage{enumitem}
\usepackage{enumerate}
\usepackage{algorithmic}

\def\ben{\begin{eqnarray*}}
\def\een{\end{eqnarray*}}

\def  \codebook{C}

\def  \H5G{H_\text{CA-Polar}}

\def  \wH{w_\text{H}}
\def  \wL{w_\text{L}}

\def  \zr{(z\circ r)}

\def\GML{ D }

\def\soft{ \Phi }

\def\numRLC{500 }

\begin{document}

%---------------------------------------------------------------------------------------------------------------------------

\title{ORDERED RELIABILITY BITS GUESSING RANDOM ADDITIVE NOISE DECODING}

\author{
\IEEEauthorblockN{Ken R. Duffy}
\IEEEauthorblockA{\textit{Hamilton Institute} \\
{Maynooth University, Ireland}\\
ken.duffy@mu.ie}
}

\maketitle

%------------------------------------------------------------------------------------------------------------------------------------

\begin{abstract}
Modern applications are driving demand for ultra-reliable low-latency
communications, rekindling interest in the performance of short,
high-rate error correcting codes. To that end, here we introduce a
soft-detection variant of Guessing Random Additive Noise Decoding
(GRAND) called Ordered Reliability Bits GRAND that can decode any
short, high-rate block-code. For a code of $n$ bits, it avails of
no more than $\lceil\log_2(n)\rceil$ bits of code-book-independent
quantized soft detection information per received bit to determine
an accurate decoding while retaining the original algorithm's
suitability for a highly parallelized implementation in hardware.
ORBGRAND is shown to provide similar block error performance for
codes of distinct classes (BCH, CA-Polar and RLC) with low complexity,
while providing better block error rate performance than CA-SCL, a
state of the art soft detection CA-Polar decoder.
\end{abstract}

\begin{IEEEkeywords}
URLLC, GRAND, Soft Decision, Quantization
\end{IEEEkeywords}

%------------------------------------------------------------------------------------------------------------------------------------
%\vspace{-0.12in}
\section{Introduction}

Maximum Likelihood (ML) decoding 
%of error correction codes 
has been
known to be optimally accurate for uniform sources since Shannon's
pioneering work \cite{Shannon48}, which also established that the
highest code-rates that a channel can support are achieved when the
code is long. Since 1978, however, it has been known that ML decoding
of linear codes is an NP-complete problem~\cite{berlekamp1978inherent}.
Taken together, these results have driven the practical paradigm
of co-designing restricted classes of linear code-books in tandem
with code-specific decoding methods that exploit the 
code-structure to enable computationally efficient approximate-ML
decoding. For example, Bose-Chaudhuri-Hocquenghem (BCH) codes with
hard detection Berlekamp-Massey decoding
\cite{berlekamp1968algebraic,massey1969shift}, or CRC-Assisted Polar
(CA-Polar) codes
with soft detection CRC-Assisted Successive
Cancellation List (CA-SCL) decoding
~\cite{niu2012crc,tal2015list,balatsoukas2015llr,leonardon2019fast}.

Recent applications, including augmented and virtual reality,
vehicle-to-vehicle communications, the Internet of Things, and
machine-type communications, have driven demand for Ultra-Reliable
Low-Latency Communication (URLLC)
\cite{durisi2016toward,she2017radio,chen2018ultra,parvez2018survey,medard20205}.
As realizing these technologies requires short, high-rate codes, the
complexity issues associated with long codes will be vacated in
URLLC and so it offers the opportunity to revisit the possibility of
creating high-accuracy universal decoders that are suitable for
hardware implementation. In turn, the development of practical
universal decoders opens up a massively larger palette of potential
code-books for decoding with a single algorithmic instantiation.

\begin{figure}
\hrule
\noindent
\begin{algorithmic}
\STATE {\bf Inputs}: Code-book membership function $\codebook:\{0,1\}^n\mapsto\{0,1\}$; 
demodulated bits $y^n$; optional information $\soft$. 
\STATE {\bf Output}: Decoding $c^{*,n}$, soft output $\GML$
\STATE $d\leftarrow 0$, $\GML\leftarrow 0$.
\WHILE{d=0}
\STATE $z^n\leftarrow$ next most likely binary noise effect sequence (which may depend on $\soft$)
\STATE $\GML\leftarrow\GML+1$ 
 \IF{$\codebook(y^n\ominus z^n) = 1$}
\STATE  $c^{*,n}\leftarrow y^n\ominus z^n$
\STATE  $d\leftarrow1$
\STATE{\bf return} $c^{*,n}$, $d$, $\GML$
\ENDIF
\ENDWHILE
\STATE
\STATE
\hrule
\end{algorithmic}
\caption{Guessing Random Additive Noise Decoding. Inputs:
a demodulated channel output $y^n$; a code-book membership
function such that $\codebook(y^n)=1$ if and only if $y^n$ is in
the code-book; and optional statistical noise characteristics or soft
information, $\soft$. Output: decoded element $c^{*,n}$;
and the number of code-book queries made, $\GML$, 
a measure of confidence in the decoding.}
\label{alg:pseudo-code}
\vspace{-0.5cm}
\end{figure}

Universal soft detection approximate-ML decoders for binary linear
codes have been proposed that work on a list-decoding principle
\cite{dorsch1974decoding,
FL95,gazelle1997reliability,VF04,wu2006soft,baldi2016use}.  In the
Orders Statistics Decoding approach, rather than compute the
conditional likelihood of the received signal for all members of
the code-book, selecting the highest likelihood one as the decoding,
instead the computation is done for a restricted list of code-words
that is hoped to contain the transmitted one. The algorithm permutes
the columns of the parity check matrix in a manner that depends on
the received signal and Gaussian elimination is performed to rewrite
the code in systematic format -- subject to checking to ensure a basis
is identified -- and the real-valued computation of conditional
probabilities, all of which make it a challenge to implement
in hardware.

A recent alternate approach that lends itself to implementation in
circuits is Guessing Random Additive Noise Decoding (GRAND)
\cite{Duffy18,Duffy19}. It is a universal ML decoding algorithm
that is suitable for use with any short, high-rate, block-code
construction in any field, including unstructured code-books stored
in a dictionary. All GRAND algorithms work by taking the demodulated
received sequence and querying if it is in the code-book. If it is
in the code-book, it is the decoding. If it is not in the code-book,
then the most likely non-zero binary noise sequence is subtracted
from the received sequence and what remains is queried for membership
of the code-book. This process proceeds until an element of the
code-book is identified, which is the decoding. Pseudo-code for
GRAND can be found in Fig. \ref{alg:pseudo-code}. Where GRAND
algorithmic variants differ is in their order of querying putative
noise effects. When putative noise effects are ordered in decreasing
likelihood from a noise model matched to the channel, it
provably produces an ML decoding even for channels with memory in
the absence of interleaving \cite{Duffy19, An20, solomon20}.  The
original algorithm assumed the decoder obtained only hard decision
demodulated symbols from the receiver. The simplicity of its
operation and the evident parallelizability of code-book
querying has resulted in the proposal of efficient circuit
implementations \cite{abbas2020high}.

Incorporating per-realization soft detection information into
decoding decisions is known to improve their accuracy
\cite{Coo88,KNH97,GS99}. Doing so, however, requires that additional
information be passed from the receiver to the decoder, which
typically necessitates its quantization for efficient transmission.
Symbol Reliability GRAND (SRGRAND) \cite{Duffy19a,Duffy19b,Duffy20}
uses the most extreme quantized soft information where one additional
bit tags each demodulated symbol as being reliably or unreliably
received.  Implementing SRGRAND in hardware retains the desirable
parallelizability and is no more challenging than doing so for
GRAND, but symbol reliability information does not exploit the full
benefit of soft information.

At the other extreme, Soft GRAND (SGRAND) \cite{solomon20} uses one
real-valued piece of soft information per demodulated symbol to
build a distinct noise-effect query order for each received signal,
resulting in soft detection ML decodings. Its execution is, however,
algorithmically involved. Using dynamic max-heap data structures
it is possible to create a semi-parallelizable software implementation,
but the design does not lend itself to realization in hardware.

Here we introduce Ordered Reliability Bits GRAND (ORBGRAND), which
aims to bridge that gap between SRGRAND and SGRAND by obtaining the
decoding accuracy of the latter in an algorithm that is entirely
parallelizable and suitable for implementation in circuits. It
avails of a code-book-independent quantization of soft information,
the decreasing rank order of the reliability of each bit in a
received block to map a fixed, pre-determined, series of putative
noise queries to their appropriate locations in the received block.
If the {\it a posteriori} bit flip probabilities match a member of
a broad parametric class of models it provides ML decodings, and
approximate-ML decodings otherwise.

\section{Ordered Reliability Bits GRAND}
\label{sec:orb}

For a block-code of length $n$ bits, all GRAND variants require a
code-book membership function $\codebook:\{0,1\}^n\mapsto\{0,1\}$
that, given a binary string of length $n$, returns $1$ if the string
is in the code-book and $0$ otherwise. For a linear code in any
field size, establishing code-book membership can be achieved with
a single matrix multiplication. For an unstructured code, checking
code-book membership can be performed by a dictionary look-up.

Assume that binary code-words, $x^n\in\{0,1\}^n$, are impacted by
independent continuous additive noise resulting in a random
received signal. Let $y^n$ denote its hard decision demodulation.
ORBGRAND provides a decoding based on the hard decision demodulation,
$y^n$, and a soft-information informed vector that records the rank
ordering of the reliability of the received bits from least reliable
to most reliable, $r^n$, which is a permutation of $(1,\ldots,n)$.
ORBGRAND does not require any further information about the received
signal or the channel.  As it attempts to identify the effect of
the noise that has impacted the communication, key to its performance
is the determination of its order of querying putative noise effect
sequences.

The algorithmic premise is to create a pre-determined rank ordered list of base noise
effect sequences, $z^{n,1}, z^{n,2},\ldots$ with the assumption
that the first bit is the least reliable, the second bit is the
second least reliable, and so forth. For a given received block
$y^n$ with rank ordered bit reliabilities described by the permutation
$r^n$, ORBGRAND maps bits in these base sequences to the correct
location via the permutation encoded in $r^n$. That is, ORBGRAND
queries the sequences $\zr^{n,1}, \zr^{n,2},\ldots$, where $\zr^{n,i}
= (z^i_{r_1},\ldots,z^i_{r_n})$.  Thus to complete the description
of the algorithm, we need only determine the ordered base sequences
$z^{n,1}, z^{n,2},\ldots$ that would be used if $r^n=(1,2,\ldots,n)$
corresponding to bits in the received block having reliability that
increases with index.

If only hard detection information was provided by the receiver and
the channel was assumed to be a binary symmetric channel, to
generate putative noise sequences in order of increasing likelihood
it is sufficient to order them in increasing Hamming weight breaking
ties arbitrarily: if $\wH(z^{n,i})<\wH(z^{n,j})$, where $\wH(z^n)=
\sum_{k=1}^n z^n_k$ is the number of ones in the sequence, $z^{n,i}$
would be queried before $z^{n,j}$.  Akin to that, the observation
underlying ORBGRAND is that if the rank ordered {\it a posteriori}
reliability of bits follow a function from a broad parametric class
described in Section \ref{sec:orb}, to rank order putative noise
sequences in increasing likelihood, it is sufficient to order them
by what we call the {\it Logistic Weight}:
\begin{align*}
\wL(z^n)= \sum_{k=1}^n k 1_{\{z^n_k=1\}}. 
\end{align*}
Thus a putative noise effect sequence in increasing order of bit reliability
is assigned a weight that is not the number of bits flipped, but
the sum of the indices of flipped bits.

\begin{figure}[h]
\begin{center}
\includegraphics[width=0.38\textwidth]{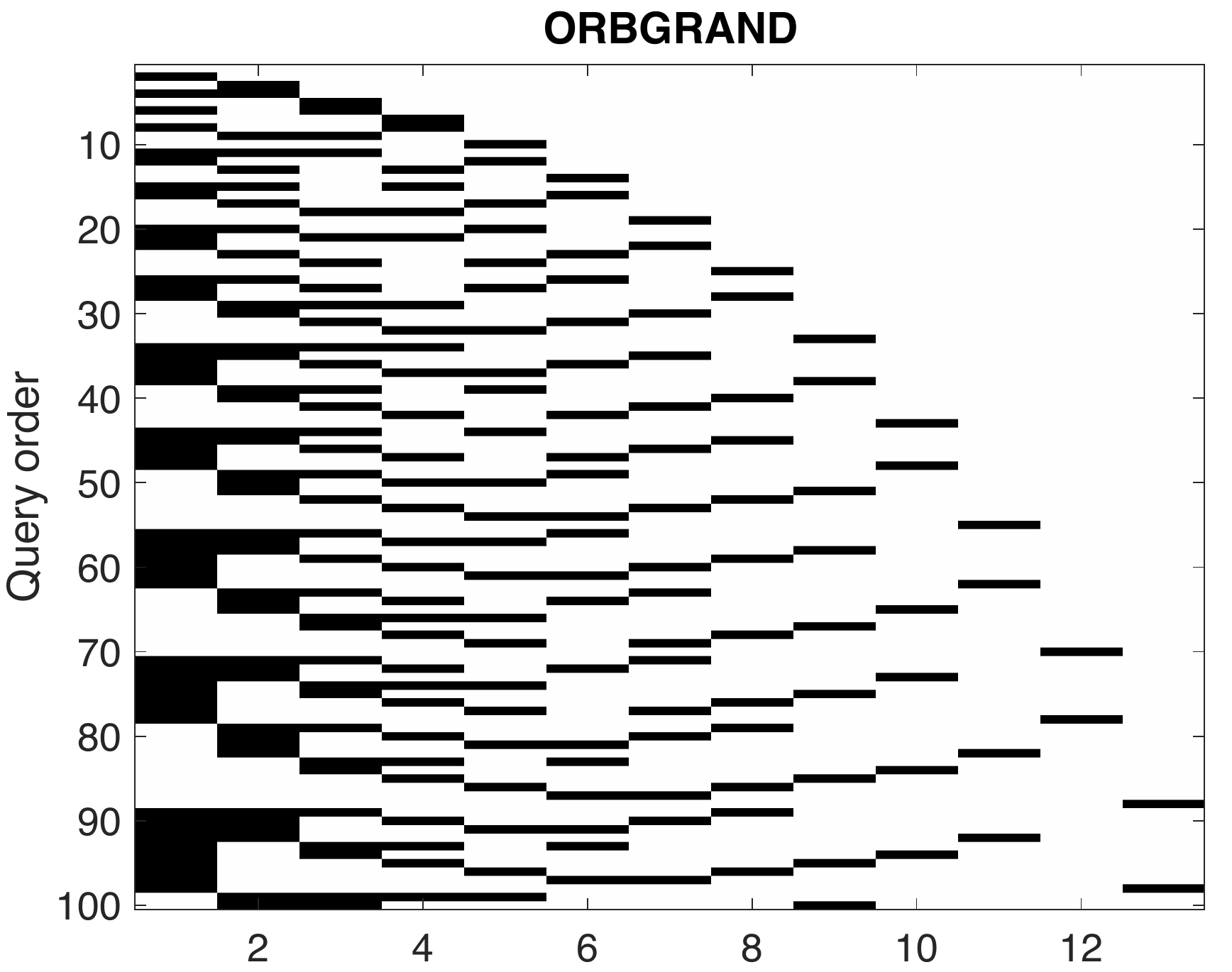}
\end{center}
\caption{First 100 ORBGRAND noise effect queries where bit positions are
in decreasing order of {\it a posteriori} bit flip likelihood. Each
row is a noise sequence with white being no bit
flip and black corresponding to a bit flip. 
}
\label{fig:2}
\vspace{-0.5cm}
\end{figure}

In the language of combinatorics, to determine the base sequences
$\{z^{n,i}\}$ requires a method to generate all integer partitions
of $\wL=0,1,2,\ldots,n(n+1)/2$ with unique parts and no part being
larger than $n$.  Efficient iterative algorithms have been described
in the literature \cite{Rasmussen1995} that can generate these
either in real-time or offline for storage, and the upper panel of
Fig.~\ref{fig:2} illustrates the first $100$ ORBGRAND queries, with
bit positions assumed to be in decreasing order of reliability. The
first sequence corresponds to no bits being flipped, which has
$\wL=0$. The second query corresponds to only the most unreliable
bit being flipped, having $\wL=1$. The third corresponds to only
the second most unreliable bit being flipped, which has $\wL=2$.
The next query is either the noise effect where only the third most
unreliable bit is flipped, which has $\wL=3$, or the one where first
and second most unreliable bits are flipped, which also has $\wL=3$,
and so this tie is broken arbitrarily. The ordering proceeds in
that fashion. While this fully describes the algorithm, it does not
justify why it works.

For a block of length $n$ bits and one use of the channel, assume
that the {\it a posteriori} bit flip probabilities, which are all
in $[0,1/2)$, are $A_1,\ldots,A_n$
and that, rank ordered from highest (i.e. least reliable) to lowest
they are $A_{1,n}\geq A_{2,n}\geq\cdots\geq A_{n,n}$.  If the latter
were consistent with a model such that
the {\it a posteriori} likelihood that the $i^\text{th}$ least reliable
hard demodulated bit $i$ was flipped was
\begin{align}
A_{n,i} = \frac{2^{-\beta i}}{1+2^{-\beta i}}
\label{eq:logistic}
\end{align}
for some $\beta>0$, then the probability of the base putative error sequence
$z^n$
that has only the bits $i_1,\cdots,i_K$ flipped would be
\begin{align*}
\prod_{i=1}^n (1-A_{n,i}) \prod_{j=1}^K \frac{A_{n,i_j}}{1-A_{n,i_j}} 
\propto \prod_{j=1}^K \frac{A_{n,i_j}}{1-A_{n,i_j}}
%=2^{-\beta \sum_{j=1}^K i_j}
=2^{-\beta \wL(z^n)}.  
\end{align*}
Thus, in terms of noise sequence likelihood, to compare the rank-order
of two base putative error sequences, one need only evaluate their
logistic weights, $\wL(z^n)=\sum_{j=1}^K i_j$, and order those.
As a result, if for a given realization Eq. \eqref{eq:logistic}
holds for any $\beta$, then ORBGRAND provides a soft ML decoding.

\begin{figure}[h]
\begin{center}
\includegraphics[width=0.38\textwidth]{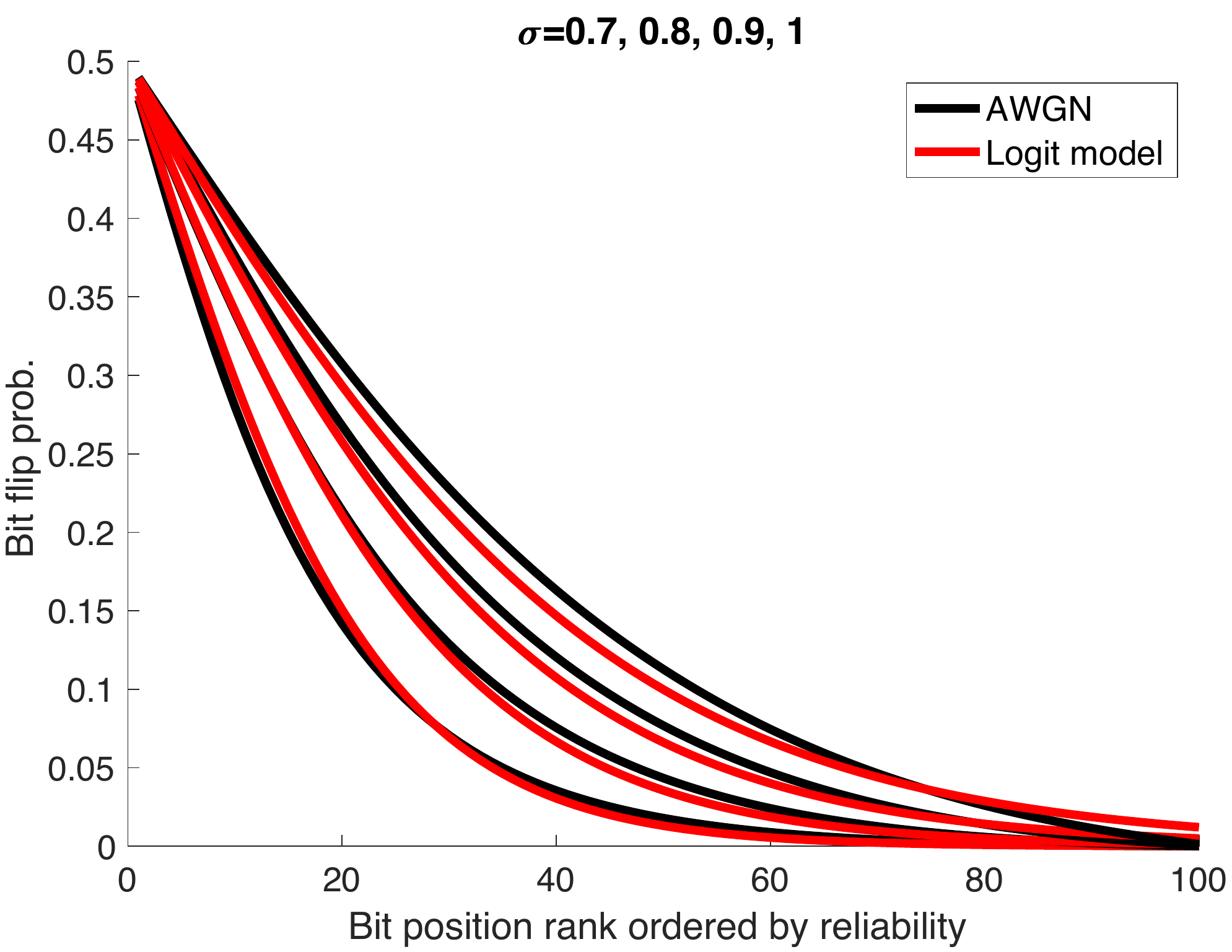}
\end{center}
\caption{Comparing {\it a posteriori} rank ordered bit flip
probabilities from a BPSK AWGN channel and the class of functions
used by ORBGRAND.}
\label{fig:4}
\vspace{-0.2cm}
\end{figure}
The richness of the class of functions in Eq. \eqref{eq:logistic}
is central to ORBGRAND's universality. Comparisons of the rank
ordered {\it a posteriori} likelihoods of bit flip from an AWGN channel
and the best-fit $\beta$ for the logistic model as determined by
regression is shown in Fig.  \ref{fig:4}, where good agreement is
found. Note that the crucial matter is not how perfect the fit is,
but whether it is good enough to produce rank-ordered noise sequences
so that the guessing order is approximately ML.

\section{Performance evaluation}
\label{sec:perf}

\begin{figure}[h]
\begin{center}
\includegraphics[width=0.38\textwidth]{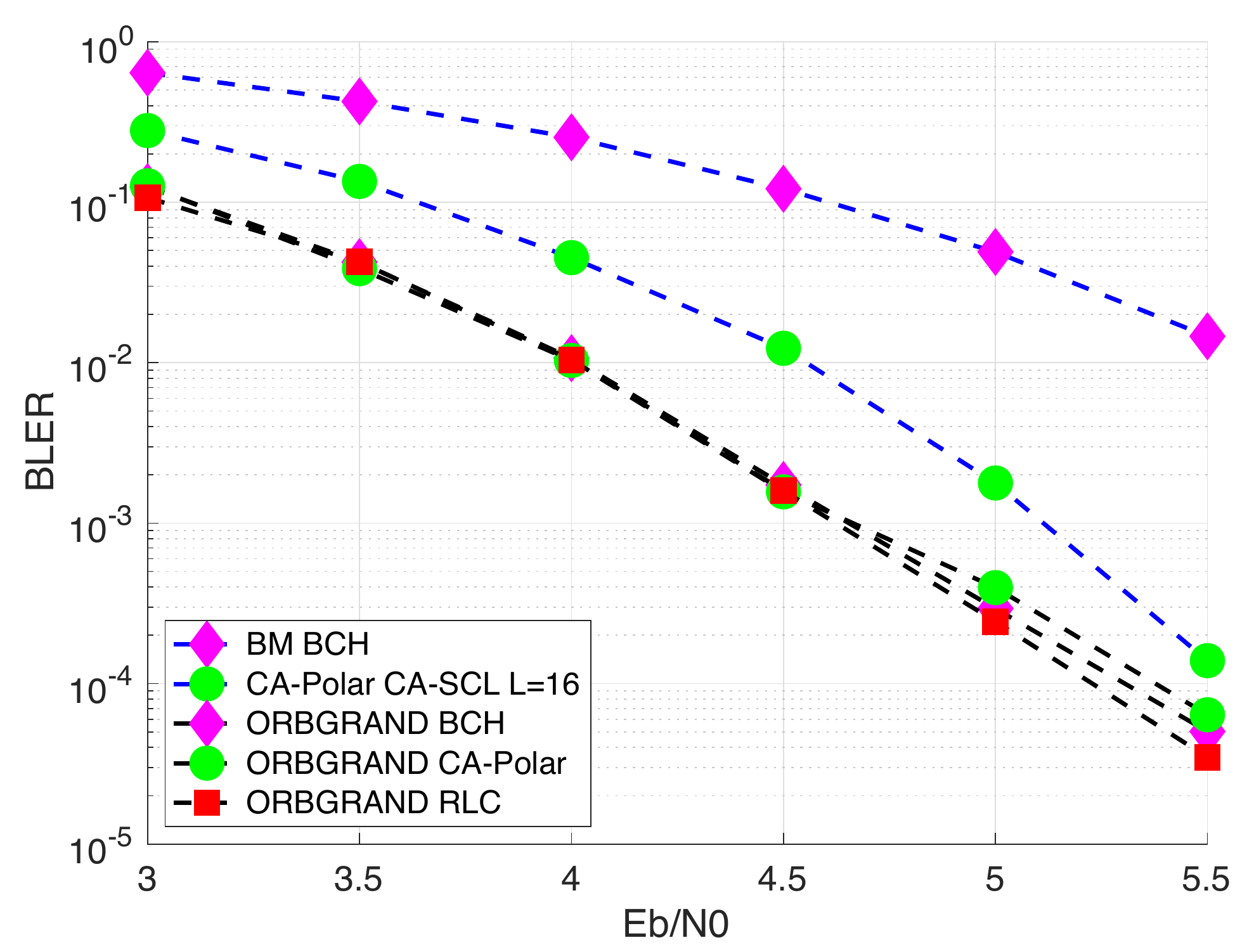}
\includegraphics[width=0.38\textwidth]{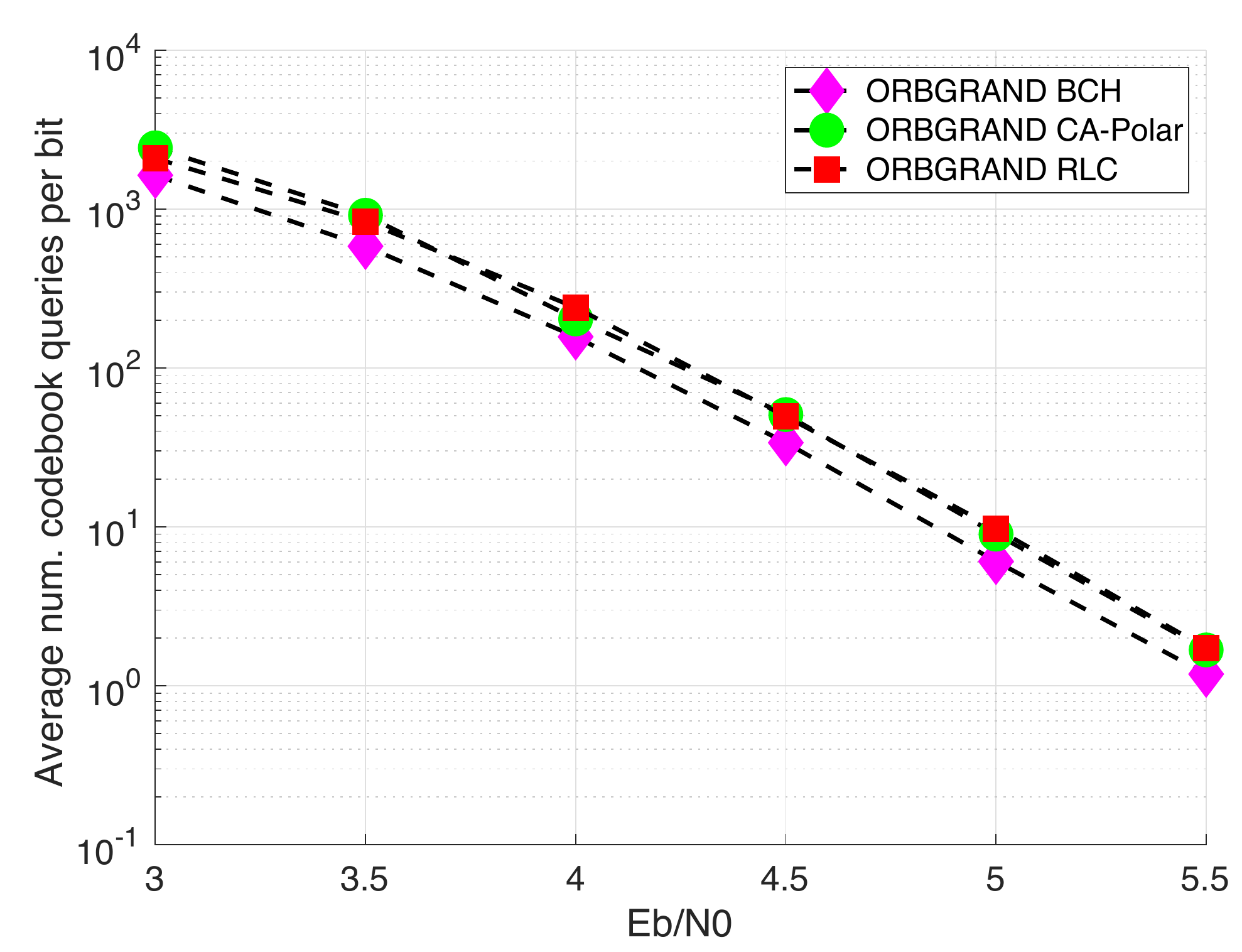}
\end{center}
\caption{Rate 0.83 BCH(127,106), and rate 0.82 CA-Polar
and RLC(128,105). Decoding BCH with hard detection Berlekamp-Massey,
CA-Polar with soft detection CA-SCL (list size $L=16$) and all three
with ORBGRAND.  Upper panel: BLER vs Eb/N0. Lower panel: average
number of code-book queries per bit until a decoding is identified
vs Eb/N0.
}
\label{fig:1}
\vspace{-0.5cm}
\end{figure}

\begin{figure}[h]
\begin{center}
\includegraphics[width=0.38\textwidth]{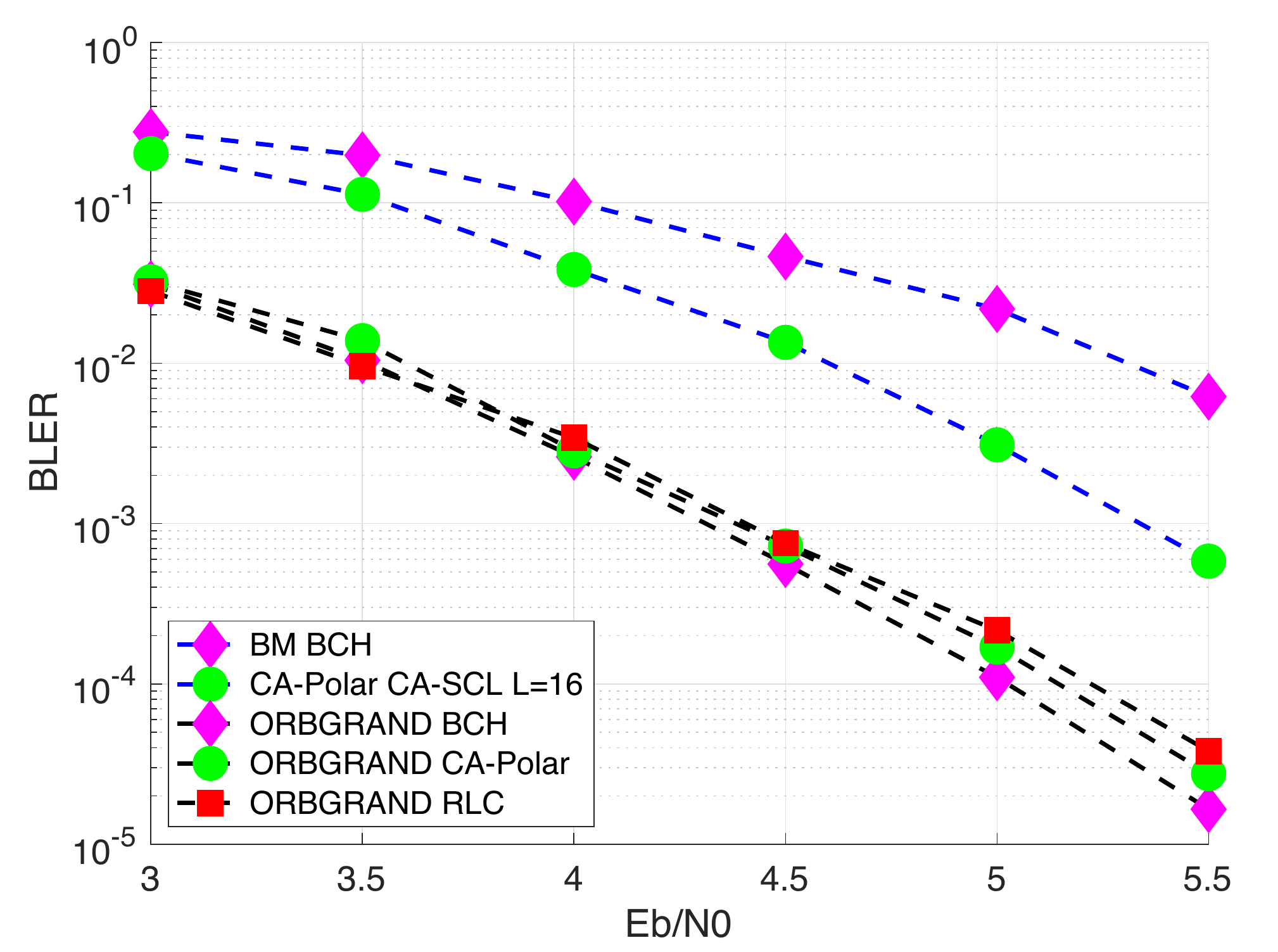}
\includegraphics[width=0.38\textwidth]{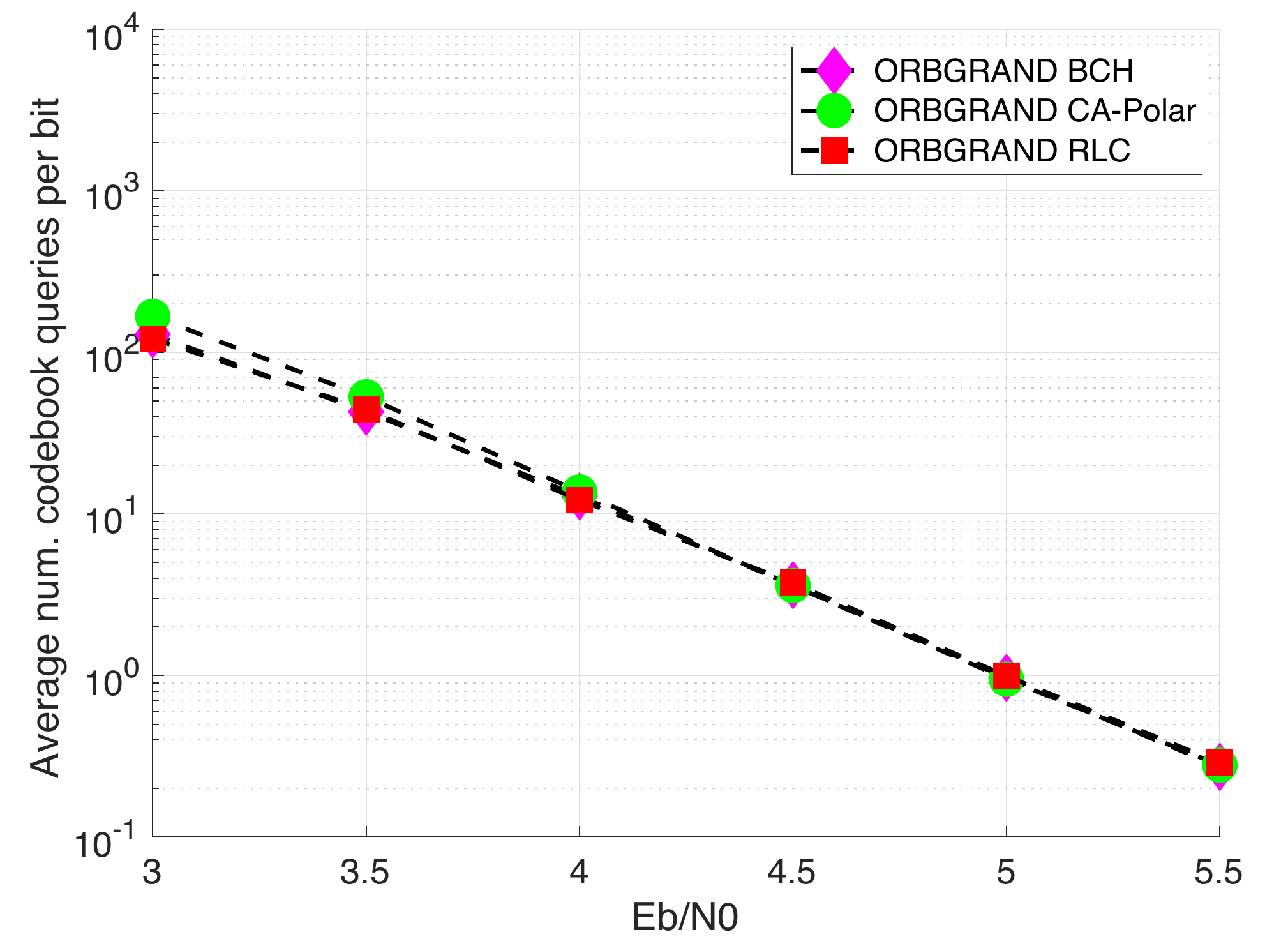}
\end{center}
\caption{As in Fig. \ref{fig:1}, but for rate $0.71$ BCH(63,45),
and rate $0.72$ CA-Polar and RLC(64,44).
}
\label{fig:5}
\vspace{-0.3cm}
\end{figure}

ORBGRAND can decode any block-code. For illustration we consider
three classes of binary linear $[n,k]$ codes: widely used BCH codes,
requiring $n$ to be a power of two minus one, for which there is
an accurate hard-detection decoder; CA-Polar codes, which will used
for all control channel communications in 5G NR and require $n$ to
be a power of two, for which there is a dedicated soft-detection
decoder, CA-SCL; and Random Linear Codes (RLCs), which includes all
linear codes and exists for all $[n,k]$ pairs. RLCs are been
theoretically known to be high-performing \cite{Gallager68,coffey1990any},
but the lack of an efficient universal decoder means their practical
potential has been little explored.

The upper panel of Fig.~\ref{fig:1} provides a block error rate
(BLER) performance comparison. Unsurprisingly, ORBGRAND's soft
detection decoding of BCH(127,106) significantly outperforms the
hard detection BM decoding. Perhaps surprisingly, ORBGRAND's decoding
of the CA-Polar(128,106) is also a half dB better than CA-SCL, the
dedicated CA-Polar decoder. This arises as CA-Polar codes are
essentially the product of two binary linear codes: a CRC and a
Polar code. In the 5G NR standard, the CRC is either 11 or 23 bits.
Here CA-SCL uses 12 Polar bits for error correction, but the 11 CRC
bits only for error detection. ORBGRAND, being a universal decoder,
uses CA-Polar code as a single code and makes use of all 23 bits
for error correction.

Also shown is the performance of RLCs of the same rate as the
CA-Polar code when decoded with ORBGRAND. A new random code is used
for each communication, offering potential additional security.
The performance of RLCs is effectively indistinguishable to that
of the BCH or CA-Polar code. This indicates that either most randomly
selected codes are as good as BCH or CA-Polar codes or there is
variability, in which case there are individual codes that have
better performance than the CA-Polar codes that will be used in 5G
NR.  Note that while there are 35 BCH codes with length 127 or
smaller, and, when an 11 bit CRC is employed, 196 CA-Polar codes
with length 128 or smaller, there are $2^{8256}$ codes available
of length 128 or smaller and ORBGRAND can decode all of them.

The lower panel of Fig.~\ref{fig:1} reports the average number of
code-book queries per bit until a decoding is found by ORBGRAND,
serving as a proxy for algorithmic complexity, though code-book
queries would be parallelized in any hardware so that many are made
per cycle \cite{abbas2020high}. As expected from theoretical
understanding of GRAND algorithms \cite{Duffy19, Duffy19b}, complexity
is essentially code-book independent. It is minimal, with about
$10^2$ queries per bit or less for typical operational settings
where a BLER of $10^{-3}$ or lower is desired.

As in Fig.~\ref{fig:1}, Fig.~\ref{fig:5} provides a performance
evaluation of ORBGRAND when used with shorter $n=64$ bit codes.
Again, by making use of the 11 bit CRC for error correction rather
than just detection, ORBGRAND sees a 1 dB gain in performance over
CA-SCL and its BLER performance is essentially code-book independent.
For these shorter codes, complexity is reduced further with
an average of less than 5 code-book queries per bit in standard
operational regimes.

\begin{figure}[h]
\begin{center}
\includegraphics[width=0.38\textwidth]{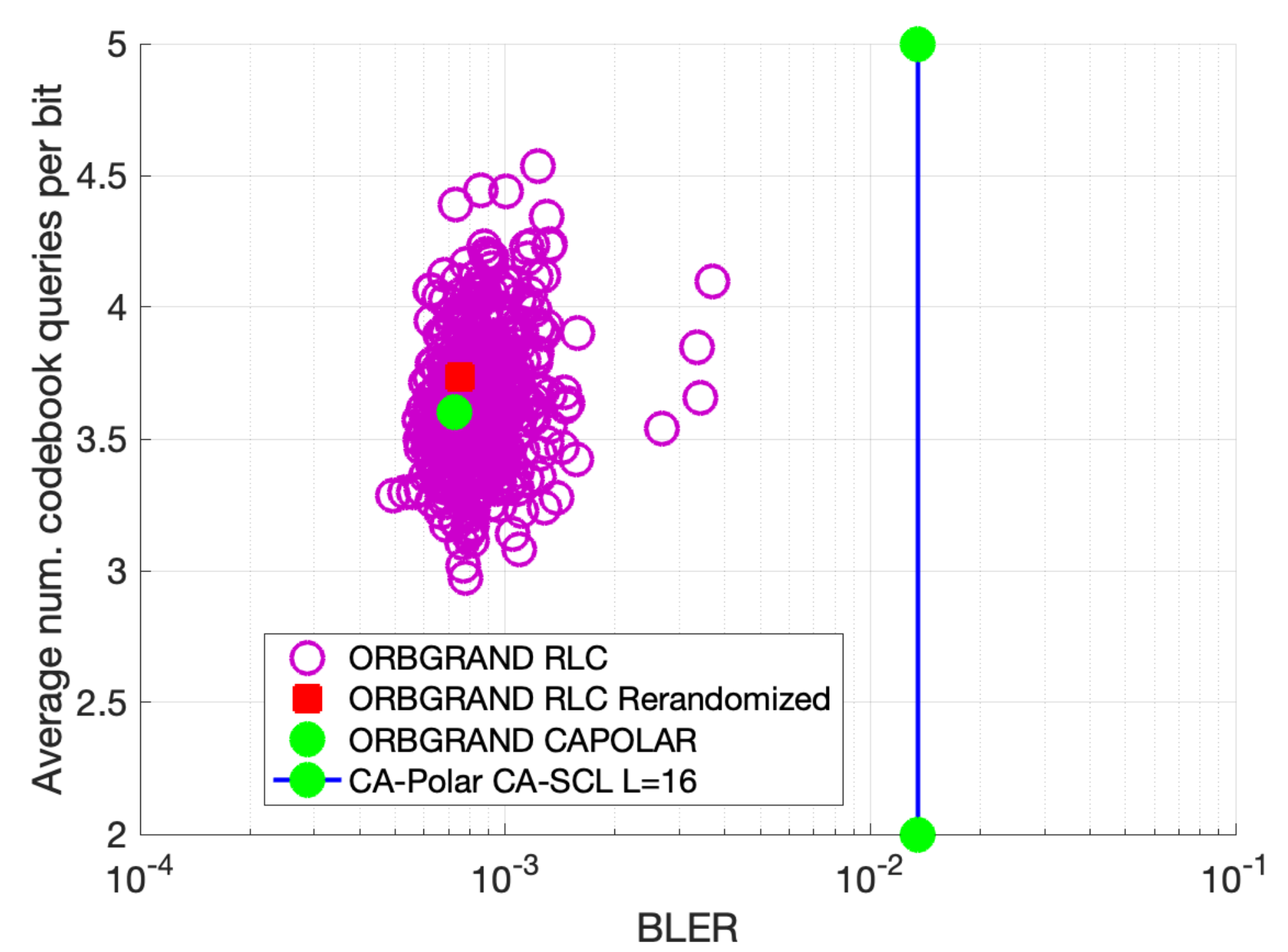}
\end{center}
\caption{BLER vs average number of code-book queries per bit 
for ORBGRAND with rate 0.71 [64,46] codes at Eb/N0=4.5 dB.
Each purple circle is one of \numRLC RLCs. The red square is 
for RLCs that are re-randomization at each communication. The green
circle is the CA-Polar code.  The vertical
line is CA-SCL with a list size
of 16, for which the y-value has no meaning.}
\label{fig:6}
\vspace{-0.4cm}
\end{figure}

The data in Figs \ref{fig:1} and \ref{fig:5} suggest that RLCs offer
similar BLER performance to well-known structured code-books. This
is further substantiated in Fig. \ref{fig:6} which, for an energy
per transmitted bit of 4.5 dB, plots BLER versus the average number
of ORBGRAND code-book queries per transmitted bit. Shown is the
performance of \numRLC individual codes, when compared with the
re-randomized RLC, and the CA-Polar code to be used in 5G NR. 22\%
of the RLCs provide better BLER than the CA-Polar code, and 50\%
give BLER that is within $10^-4$ of it. For reference the CA-SCL
BLER result is also plotted.

\section{Summary}
We have introduced ORBGRAND, a universal block decoding algorithm
that uses a code-book-independent quantization of soft information
to inform its decoding. It is suitable for hardware implementation,
and results indicate it can provide better BLER performance for
short 5G NR CA-Polar codes than a state-of-the-art soft detection
decoder.  It is also shown that Random Linear Codes offer essentially
the same performance as highly structured BCH and CA-Polar codes,
opening up new application possibilities.

%\bibliographystyle{IEEEtran}
%\bibliography{ORBGRAND,SRGRAND}


\end{document}